\documentclass[a4paper,10pt]{article}
\usepackage{stmaryrd}
\usepackage{amsfonts}
\usepackage{bbm}
\usepackage{amscd}
\usepackage{mathrsfs}
\usepackage{latexsym,amssymb,amsmath,amscd,amscd,amsthm,amsxtra,xypic}
\usepackage[dvips]{graphicx}
\usepackage[utf8]{inputenc}
\usepackage[T1]{fontenc}
\usepackage{lmodern}
\usepackage{amssymb}
\usepackage[all]{xy}
\usepackage{nicefrac,mathtools,enumitem}
\usepackage{microtype}

\newtheorem{thm}{Theorem}[section]
\newtheorem{lem}[thm]{Lemma}
\newtheorem{cor}[thm]{Corollary}
\newtheorem{pro}[thm]{Proposition}

\newtheorem{rmk}[thm]{Remark}
\newtheorem{defi}[thm]{Definition}

\newcommand{\be }{\begin{equation}}
\newcommand{\ee }{\end{equation}}

\newcommand{\pf}{\noindent{\bf Proof.}\ }

\newcommand{\huaG}{\mathcal{G}}

\newcommand{\huaV}{\mathcal{V}}

\newcommand{\huaT}{\mathcal{T}}

\newcommand{\CWM}{C^{\infty}(M)}

\newcommand{\frkf}{\mathfrak f}
\newcommand{\frkg}{\mathfrak g}
\newcommand{\frkh}{\mathfrak h}

\newcommand{\frkX}{\mathfrak X}

\def\qed{\hfill ~\vrule height6pt width6pt depth0pt}

\newcommand{\half}{\frac{1}{2}}

\newcommand{\pair}[1]{\left\langle #1\right\rangle}

\newcommand{\Courant}[1]{\left\llbracket  #1\right\rrbracket }
\newcommand{\Dorfman}[1]{\{ #1\}}

\newcommand{\Id}{\rm{Id}}

\newcommand{\br}[1]{   [ \cdot,    \cdot  ]   }
\newcommand{\id}{\mathbbm{i}}

\newcommand{\dM}{\mathrm{d}}

\newcommand{\Hom}{\mathrm{Hom}}

\newcommand{\gl}{\mathfrak {gl}}

\newcommand{\Ker}{\mathrm{Ker}}

\newcommand{\End}{\mathrm{End}}

\newcommand{\ve}{\mathrm{v}}
\newcommand{\Vect}{\mathrm{Vect}}
\newcommand{\sgn}{\mathrm{sgn}}
\newcommand{\Ksgn}{\mathrm{Ksgn}}

\newcommand{\V}{\mathbb{V}}

\begin{document}
\title{
{Leibniz $2$-algebras and twisted Courant algebroids
\thanks
 {
Research supported by NSFC (10871007,11026046), Doctoral Fund. of
MEC (20090001110006,20100061120096) and "the Fundamental Research
Funds for the Central Universities" (200903294).
 }
} }
\author{Yunhe Sheng  \\
Department of Mathematics, Jilin University,\\
 Changchun 130012,  China
\\\vspace{3mm}
email: shengyh@jlu.edu.cn\\
Zhangju Liu\\
Department of Mathematics and LMAM, Peking University, \\Beijing
100871, China\\\vspace{3mm} email: liuzj@pku.edu.cn}
\date{}
\maketitle

\begin{abstract}
  In this paper, we give the categorification of Leibniz algebras,
  which is equivalent to 2-term sh Leibniz algebras.
They reveal the algebraic structure of omni-Lie 2-algebras
introduced in \cite{omniLie2} as well as twisted Courant algebroids
by closed $4$-forms introduced in \cite{4form}.
  We also prove that
 Dirac structures of twisted Courant algebroids give rise to 2-term
$L_\infty$-algebras and  geometric structures behind them are
exactly $H$-twisted Lie algebroids introduced in \cite{Grutzmann}.
\end{abstract}


\section{Introduction}
Recently, people have payed more attention to higher categorical
structures by reasons in both mathematics and physics. One way to
provide higher categorical structures is by categorifying existing
mathematical concepts. One of the simplest higher structures is a
$2$-vector space, which is the  categorification of a vector space.
If we further put a compatible Lie algebra structure on a $2$-vector
space, then we obtain a Lie $2$-algebra
\cite{baez:2algebras,RoytenbergWeakLie2}. The Jacobi identity is
replaced by a natural transformation, called the Jacobiator, which
also satisfies some coherence laws of its own. Recently, the
relation among higher categorical structures and multisymplectic
structures, Courant algebroids, and Dirac structures are  studied in
\cite{baezrogers,rogers,zambon}.

A $2$-vector space is equivalent to a $2$-term complex of vector
spaces. A Lie $2$-algebra is equivalent to a $2$-term
$L_\infty$-algebra. $L_\infty$-algebras,  sometimes  called strongly
homotopy (sh) Lie algebras,  were introduced in
\cite{Stasheff1,stasheff2} as a model for ``Lie algebras that
satisfy Jacobi identity up to all higher homotopies''. The notion of
Leibniz algebras was introduced by Loday \cite{Loday}, which is a
generalization of Lie algebras. Their crossed modules were also
introduced in \cite{Loday and Pirashvili} to study the cohomology of
Leibniz algebras. As a  model for ``Leibniz algebras that satisfy
Jacobi identity up to all higher homotopies'', Ammar and Poncin
introduced the notion of strongly homotopy Leibniz algebra, or
$Lod_\infty$-algebra in \cite{ammardefiLeibnizalgebra}, which is
further studied by Uchino in \cite{UchinoshL}.

 Courant algebroid was introduced in \cite{LWXmani} to study the
double of Lie bialgebroids. Equivalent definition was given by
Roytenberg in \cite{roytenbergshl}. Courant algebroids have been
widely studied because of their applications in both mathematics and
physics. Roytenberg proved that every Courant algebroid give rise to
an $L_\infty$-algebra \cite{roytenbergshl}. The $L_\infty$-algebra
associated to the standard Courant algebroid $TM \oplus T^*M$ is a
semidirect product of a Lie algebra with a representation up to
homotopy \cite{SZsemidirectproduct,SZintegration}. Recently, Hansen
and Strobl introduced the notion of twisted Courant algebroids by
closed $4$-forms in \cite{4form}, which arise from the study of
three dimensional sigma models with Wess-Zumino term. Similar
structures were studied in \cite{preCourant,GrutzmannC,Vai05}. In
general, if one studies generalized geometry, this $4$-form will
arise naturally as background \cite{Hull}. A closed $4$-form is also
used to construct a bundle 2-gerbe in \cite{higher}.

In \cite{Pontryagin class} and  \cite{regular CA},  the {\em first
Pontryagin class} was introduced for a quadratic Lie algebroid,
which is a closed $4$-form  as an obstruction of the extension to a
Courant algebroid. It will be seen  that a quadratic Lie algebroid
with nontrivial first Pontryagin class can be realized as the ample
Lie algebroid of a regular twisted Courant algebroid, this is a
generalization of \cite[Theorem 1.10]{regular CA}.

 In this paper, we introduce the notion of Leibniz 2-algebras, which is equivalent to 2-term strongly homotopy Leibniz algebras.
 Similar to the case of Lie algebras, we prove that
there is a one-to-one correspondence between $2$-term dg Leibniz
algebras and crossed modules of Leibniz algebras. With the help of
an automorphism $\frkf$ of the $2$-term DGLA $\End(\huaV)$, where
$\huaV$ is a $2$-term complex of vector spaces, we construct a
Leibniz 2-algebra $(\End(\huaV)\oplus\huaV,l_2^\frkf,l_3^\frkf)$,
which essentially comes from omni-Lie $2$-algebra introduced in
\cite{omniLie2}. Every twisted Courant algebroid by a closed 4-form
$H$
  gives rise to a Leibniz 2-algebra.
In particular, Dirac structures of twisted Courant algebroids give
rise to $2$-term $L_\infty$-algebras. The geometric structure
underlying this $2$-term $L_\infty$-algebra is $H$-twisted Lie
algebroid introduced by Gr{\"u}tzmann in \cite{Grutzmann}. $B$-field
transformation \cite{GualtieriGeneralizedComplex} is an important
tool to provide symmetries of exact Courant algebroids. In a
$B$-field transformation, the $2$-form need to be closed.  Now for
exact twisted Courant algebroids, every $2$-form (not need to be
closed) provides an automorphism of the corresponding Leibniz
2-algebra.

The paper is organized as follows. In Section \ref{sec:pre} 
we prove that there is a one-to-one correspondence between $2$-term
dg Leibniz algebras and crossed modules of Leibniz algebras (Theorem
\ref{thm:correspondence}).  In Section \ref{sec:leibniz2} we
introduce the notion of Leibniz 2-algebra, which is the
categorification of Leibniz algebras. We show that they are
equivalent to 2-term sh Leibniz algebras. In Section
\ref{sec:example} associated to any automorphism $\frkf$ of
$\End(\huaV)$, we construct a Leibniz 2-algebra
$(\End(\huaV)\oplus\huaV,l_2^\frkf,l_3^\frkf)$. In Section
\ref{sec:tca} we show that every twisted Courant algebroid gives
rise to a Leibniz 2-algebra (Theorem \ref{thm:Leibniz2algebra}). Via
the $B$-field transformation, any $2$-form provides an automorphism
of the Leibniz 2-algebra associated to an exact twisted Courant
algebroid (Theorem \ref{thm:automorphism}). In Section
\ref{sec:Dirac}  we study Dirac structures of a twisted Courant
algebroid, it turns out that a Dirac structure of a twisted Courant
algebroid gives rise to a $2$-term $L_\infty$-algebra (i.e. a Lie
2-algebra, Theorem \ref{thm:Lie2Dirac}). We also find that the
geometric structure underlying this $2$-term $L_\infty$-algebra is
$H$-twisted Lie algebroid. At last, we consider the Dirac structure
$\huaG_\pi$, which is the graph of a bi-vector field $\pi$, and
obtain $h$-twisted Poisson structure (the 3-form $h$ is not closed)
as well as  the associated $2$-term $L_\infty$-algebra. \vspace{3mm}

{\bf Acknowledgement:} We would like to thank  P. Bouwknegt, M.
Gr{\"u}tzmann, Sen Hu, Bailing Wang and  Chenchang Zhu  for helpful
discussions and comments. We also give our special thanks to
referees for many helpful suggestions.

\section{Crossed modules of Leibniz algebras and sh Leibniz
algebras}\label{sec:pre}

 A Leibniz algebra $\frkg$ is an
$R$-module, where $R$ is a commutative ring, endowed with a linear
map $[\cdot,\cdot]_\frkg:\frkg\otimes\frkg\longrightarrow\frkg$
satisfying
$$
[g_1,[g_2,g_3]_\frkg]_\frkg=[[g_1,g_2]_\frkg,g_3]_\frkg+[g_2,[g_1,g_3]_\frkg]_\frkg,\quad
\forall~g_1,g_2,g_3\in \frkg.
$$

This is in fact a left Leibniz algebra. In this paper, we only
consider left Leibniz algebras.

 Recall that a representation of the Leibniz algebra
$(\frkg,[\cdot,\cdot]_\frkg)$ is an $R$-module $V$ equipped with,
respectively, left and right actions  of $\frkg$ on $V$,
$$[\cdot,\cdot]:\frkg\otimes V\longrightarrow V,\quad [\cdot,\cdot]:V\otimes\frkg \longrightarrow V,$$
such that for any $g_1,g_2\in\frkg$, the following equalities hold:
\begin{equation}\label{condition of rep}
l_{[g_1,g_2]}=[l_{g_1},l_{g_2}],\quad
r_{[g_1,g_2]}=[l_{g_1},r_{g_2}],\quad r_{g_2}\circ
l_{g_1}=-r_{g_2}\circ r_{g_1},
\end{equation} where
$l_{g_1}u=[g_1,u]$ and $r_{g_1}u=[u,g_1]$ for any $u\in V$. The
Leibniz cohomology of $\frkg$ with coefficients in $V$ is the
homology of the cochain complex
$C^k(\frkg,V)=\Hom_R(\otimes^k\frkg,V), (k\geq0)$ with the
coboundary operator $\partial:C^k(\frkg,V)\longrightarrow
C^{k+1}(\frkg,V)$ defined by
\begin{eqnarray}
\nonumber\partial
c^k(g_1,\dots,g_{k+1})&=&\sum_{i=1}^k(-1)^{i+1}l_{g_i}(c^k(g_1,\dots,\widehat{g_i},\dots,g_{k+1}))
+(-1)^{k+1}r_{g_{k+1}}(c^k(g_1,\dots,g_k))\\
\label{formulapartial}&&+\sum_{1\leq i<j\leq
k+1}(-1)^ic^k(g_1,\dots,\widehat{g_i},\dots,g_{j-1},[g_i,g_j]_\frkg,g_{j+1},\dots,g_{k+1}).
\end{eqnarray}
The fact that $\partial\circ\partial=0$ is proved in \cite{Loday and
Pirashvili}. \vspace{3mm}

The notion of strongly homotopy (sh) Leibniz algebras, or
$Lod_\infty$-algebras was first given in
\cite{ammardefiLeibnizalgebra}. See also \cite{UchinoshL} for more
details.

\begin{defi}{\rm\cite{UchinoshL}}
A sh Leibniz algebra is a graded  vector space $L=L_0\oplus
L_1\oplus\cdots$ equipped with a system $\{l_k|~1\leq k<\infty\}$ of
linear maps $l_k:\wedge^kL\longrightarrow L$ with degree
$\deg(l_k)=k-2$, where the exterior powers are interpreted in the
graded sense and the following relation  is satisfied:
\begin{eqnarray*}
&&\sum_{i+j=Const}\sum_{k\geq
j}\sum_{\sigma}(-1)^{(k+1-j)(j-1)}(-1)^{j(|x_{\sigma(1)}|+\cdots+|x_{\sigma(k-j)}|)}\sum_{\sigma}\sgn(\sigma)\Ksgn(\sigma)\\
&&l_i(x_{\sigma(1)},\dots,x_{\sigma(k-j)},l_j(x_{\sigma(k+1-j)},\dots,x_{\sigma(k)}),x_{\sigma(k+1)},\dots,x_{\sigma(i+j-1)})=0,
\end{eqnarray*}
where the summation is taken over all $(k-j,j-1)$-unshuffles and
``$\Ksgn(\sigma)$'' is the Koszul sign for a permutation $\sigma\in
S_k$, i.e. $$ x_1\wedge x_2\wedge\cdots\wedge
x_k=\Ksgn(\sigma)x_{\sigma(1)}\wedge x_{\sigma(2)}\wedge\cdots\wedge
x_{\sigma(k)}.
$$
\end{defi}

In particular, if we concentrate on the 2-term case, we can give
explicit formulas for 2-term sh Leibniz algebras as follows:

\begin{defi}\label{defi:2leibniz}
  A $2$-term sh Leibniz algebra $\huaV$ consists of the following data:
\begin{itemize}
\item[$\bullet$] a complex of vector spaces $\huaV:V_1\stackrel{\dM}{\longrightarrow}V_0,$

\item[$\bullet$] bilinear maps $l_2:V_i\times V_j\longrightarrow
V_{i+j}$, where $i+j\leq 1$,

\item[$\bullet$] a  trilinear map $l_3:V_0\times V_0\times V_0\longrightarrow
V_1$,
   \end{itemize}
   such that for any $w,x,y,z\in V_0$ and $m,n\in V_1$, the following equalities are satisfied:
\begin{itemize}
\item[$\rm(a)$] $\dM l_2(x,m)=l_2(x,\dM m),$
\item[$\rm(b)$]$\dM l_2(m,x)=l_2(\dM m,x),$
\item[$\rm(c)$]$l_2(\dM m,n)=l_2(m,\dM n),$
\item[$\rm(d)$]$\dM l_3(x,y,z)=l_2(x,l_2(y,z))-l_2(l_2(x,y),z)-l_2(y,l_2(x,z)),$
\item[$\rm(e_1)$]$ l_3(x,y,\dM m)=l_2(x,l_2(y,m))-l_2(l_2(x,y),m)-l_2(y,l_2(x,m)),$
\item[$\rm(e_2)$]$ l_3(x,\dM m,y)=l_2(x,l_2(m,y))-l_2(l_2(x,m),y)-l_2(m,l_2(x,y)),$
\item[$\rm(e_3)$]$ l_3(\dM m,x,y)=l_2(m,l_2(x,y))-l_2(l_2(m,x),y)-l_2(x,l_2(m,y)),$
\item[$\rm(f)$] the Jacobiator identity:\begin{eqnarray*}
&&l_2(w,l_3(x,y,z))-l_2(x,l_3(w,y,z))+l_2(y,l_3(w,x,z))+l_2(l_3(w,x,y),z)\\
&&-l_3(l_2(w,x),y,z)-l_3(x,l_2(w,y),z)-l_3(x,y,l_2(w,z))\\
&&+l_3(w,l_2(x,y),z)+l_3(w,y,l_2(x,z))-l_3(w,x,l_2(y,z))=0.\end{eqnarray*}
   \end{itemize}
\end{defi}
We usually denote a 2-term sh Leibniz algebra by
$(V_1\stackrel{\dM}{\longrightarrow}V_0,l_2,l_3)$, or simply by
$\huaV$.\vspace{3mm}

If $l_3=0$, we obtain the notion of {\bf $2$-term differential
graded (dg) Leibniz algebra}. If the bilinear maps $l_2$ and the
trilinear map $l_3$ are skew-symmetric, then it is a {\bf $2$-term
$L_\infty$-algebra}.

\begin{lem}\label{lem:rep1}
For a $2$-term dg Leibniz algebra
$(V_1\stackrel{\dM}{\longrightarrow}V_0,l_2,l_3)$, we have
\begin{equation}\label{eqn:rep1}
l_2(l_2(x,m),y)+l_2(l_2(m,x),y)=0,\quad\forall~x,y\in V_0,m\in V_1.
\end{equation}
\end{lem}
\pf By Condition $(e_2)$ and $(e_3)$ in Definition
\ref{defi:2leibniz}, we have
\begin{eqnarray*}
 l_2(l_2(x,m),y)+l_2(l_2(m,x),y)&=&l_2(x,l_2(m,y))-l_2(m,l_2(x,y))\\
 &&+l_2(m,l_2(x,y))-l_2(x,l_2(m,y))\\
 &=&0. \qed
\end{eqnarray*}

The notion of crossed module of Leibniz algebras was introduced by
 Loday and Pirashvili in \cite{Loday and Pirashvili}. The
more general notion of crossed module of $n$-Leibniz algebras, which
are generalizations of $n$-Lie algebras, was given by Casas,
Khmaladze and Ladra in \cite{cmL}.
\begin{defi}
 A crossed module of Leibniz algebras is a morphism of Leibniz algebras $\mu:\frkg\longrightarrow\frkh$ together with
a representation of $\frkh$ (consists of a left action and a right
action satisfying the compatibility condition \eqref{condition of
rep}) on $\frkg$  such that for any $g,
g^\prime\in\frkg,~h\in\frkh$, the following equalities hold:
\begin{eqnarray}
 \label{condition1} \mu (l_hg)&=&[h,\mu(g)]_\frkh,\quad \mu(r_hg)=[\mu(g),h]_\frkh;\\
  \label{condition2}~l_{\mu(g)}g^\prime&=&[g,g^\prime]_\frkg=r_{\mu(g^\prime)}g;\\
 \label{condition3} l_h[g,g^\prime]_\frkg&=&[l_hg,g^\prime]_\frkg+[g,l_hg^\prime]_\frkg;\\
 \label{condition4}
 r_h[g,g^\prime]_\frkg&=&[g,r_hg^\prime]_\frkg-[g^\prime,r_hg];\\
 \label{condition5} ~[l_hg+r_hg,g^\prime]_\frkg&=&0.
\end{eqnarray}
\end{defi}

\begin{rmk}
Loday and Pirashvili defined a crossed module of Leibniz algebras to
be a morphism of Leibniz algebras $\mu:\frkg\longrightarrow\frkh$
together with an action of $\frkh$ on $\frkg$ satisfying
\eqref{condition1} and \eqref{condition2} in \cite{Loday and
Pirashvili}. However, to define an action of Leibniz algebra $\frkh$
on Leibniz algebra $\frkg$, one needs six relations, which is
exactly \eqref{condition of rep}, \eqref{condition3},
\eqref{condition4}, and \eqref{condition5}.
\end{rmk}

\begin{thm}\label{thm:correspondence}
  There is a one-to-one correspondence between $2$-term dg
Leibniz algebras and crossed modules of Leibniz algebras.
\end{thm}
\pf Let $V_1\stackrel{\dM}{\longrightarrow} V_0$ be a $2$-term dg
Leibniz algebra, define $\frkg=V_1$, $\frkh=V_0$, and the following
two bracket operations on $\frkg$ and $\frkh$:
\begin{eqnarray*}
~[m,n]_{\frkg}&=&l_2(\dM m,n)=l_2( m,\dM n),\quad\forall~m,n\in
V_1;\\
~[u,v]_{\frkh}&=&l_2(u,v),\quad\forall~u,v\in V_0.
\end{eqnarray*}
It is straightforward to see that both $[\cdot,\cdot]_\frkg$ and
$[\cdot,\cdot]_\frkh$ are Leibniz brackets. Let $\mu=\dM$, by
Condition (a) in Definition \ref{defi:2leibniz}, we have
$$
\mu[m,n]_\frkg=\dM l_2(\dM m,n)=l_2(\dM m,\dM
n)=[\mu(m),\mu(n)]_\frkh,
$$
which implies that $\mu$ is a morphism of Leibniz algebras. Define
the representation of $\frkh$ on $\frkg$ by $l_2$, i.e.
$$
l_um=l_2(u,m),\quad r_um=l_2(m,u),\quad \forall~ u\in \frkh,
m\in\frkg.
$$
It is well defined. In fact, by Condition $\rm (e_1)$, we have
$$
l_{[u,v]_\frkh}=[l_u,l_v].
$$
By \eqref{eqn:rep1}, we have
$$
r_vr_u+r_vl_u=0.
$$
Now by Condition $\rm (e_3)$, we have
$$
 r_{[u,v]_\frkh}=[l_u,r_v],
$$
which implies that \eqref{condition of rep} holds.

By Conditions (a)-(c), we have \eqref{condition1} and
\eqref{condition2}. By Condition $\rm (e_1)$ and (a), we have
\begin{eqnarray*}
  l_u[m,n]_\frkg&=&l_2(u,l_2(\dM m,n))\\
  &=&l_2(l_2(u,\dM m),n)+l_2(\dM m,l_2(u,n))\\
  &=&l_2(\dM l_2(u, m),n)+l_2(\dM m,l_2(u,n))\\
  &=&[l_um,n]_\frkg+[m,l_un]_\frkg,
\end{eqnarray*}
which yields \eqref{condition3}. By Condition $\rm (e_2)$, we have
\begin{eqnarray*}
  r_u[m,n]_\frkg&=&l_2(l_2(\dM m,n),u)\\
  &=&l_2(\dM m,l_2(n,u))-l_2(n,l_2(\dM m,u))\\
  &=&[m,r_un]_\frkg-[n,r_um]_\frkg,
\end{eqnarray*}
which implies that \eqref{condition4} holds. By \eqref{eqn:rep1}, we
have
 \begin{eqnarray*}
[ r_um+l_um,n]_\frkg&=&l_2(l_2(u,m)+l_2(m,u),\dM n)=0.
\end{eqnarray*}
Thus we get \eqref{condition5}. Therefore, we obtain a crossed
module of Leibniz algebras.

Conversely,  a crossed module of Leibniz algebras
 gives rise to a 2-term dg Leibniz algebra with $\dM=\mu$,
$V_1=\frkg$ and $V_0=\frkh$, where the brackets are given by:
\begin{eqnarray*}
~ l_2(u,v)&\triangleq&[u,v]_{\frkh_0},\quad \forall
~u,v\in\frkh;\\
~l_2(u,m)&\triangleq&l_um,\quad\forall~ m\in\frkh_1;\\
~l_2(m,u)&\triangleq&r_um.
\end{eqnarray*}
The crossed module conditions give various conditions for $2$-term
dg Leibniz algebras. We omit details.\qed

\begin{defi}
   A  $2$-term sh Leibniz algebra $(V_1\stackrel{\dM}{\longrightarrow}V_0,l_2,l_3)$ is called skeletal if $\dM=0$.
\end{defi}
 Skeletal $2$-term sh Leibniz algebras can be classified by the
 third cohomology of Leibniz algebras.

\begin{pro}
  There is a one-to-one correspondence between skeletal $2$-term sh Leibniz algebras $(V_1\stackrel{0}{\longrightarrow}V_0,l_2,l_3)$ and
  quadruples $(\frkg,V,\rho,\phi)$, where $\frkg$ is a Leibniz
  algebra, $V$ is a vector space, $\rho$ is a representation of
  $\frkg$ on $V$, $\phi$ is $3$-cocycle on $\frkg$ with coefficients in
  $V$.
\end{pro}
\pf For a skeletal $2$-term sh Leibniz algebra
$(V_1\stackrel{0}{\longrightarrow}V_0,l_2,l_3)$, by
  Condition $(d)$ in Definition \ref{defi:2leibniz}, $V_0$ is a
  Leibniz algebra. By Condition $(e_1), (e_3)$ in Definition \ref{defi:2leibniz} and \eqref{eqn:rep1}
  in Lemma \ref{lem:rep1}, we get that $l_2:V_i\times V_j\longrightarrow V_1  (i+j=1)$  gives rise to a
  representation of Leibniz algebra $V_0$ on $V_1$. Now Condition
  $(f)$ means that $\partial l_3(w,x,y,z)=0$ by Formula
  \eqref{formulapartial}.

The converse part is also straightforward, this completes the proof.
\qed

\begin{defi}\label{defi:Leibniz morphism}
 Let $\huaV$ and $\huaV^\prime$ be  $2$-term sh Leibniz algebras, a morphism $\frkf$
 from $\huaV$ to $\huaV^\prime$ consists of
\begin{itemize}
  \item[$\bullet$] linear maps $f_0:V_0\longrightarrow V_0^\prime$
  and $f_1:V_1\longrightarrow V_1^\prime$ commuting with the
  differential, i.e.
  $$
f_0\circ \dM=\dM^\prime\circ f_1;
  $$
  \item[$\bullet$] a bilinear map $f_2:V_0\times V_0\longrightarrow
  V_1^\prime$,
\end{itemize}
  such that  for all $x,y,z\in L_0,~m\in
L_1$, we have
\begin{equation}\label{eqn:DGLA morphism c 1}\left\{\begin{array}{rll}
l_2^\prime(f_0(x),f_0(y))-f_0l_2(x,y)&=&\dM^\prime
f_2(x,y),\\
l_2^\prime(f_0(x),f_1(m))-f_1l_2(x,m)&=&f_2(x,\dM
m),\\
l_2^\prime(f_1(m),f_0(x))-f_1l_2(m,x)&=&f_2(\dM
m,x),\end{array}\right.
\end{equation}
and
\begin{eqnarray}
\nonumber&&f_1(l_3(x,y,z))+l_2^\prime(f_0(x),f_2(y,z))-l_2^\prime(f_0(y),f_2(x,z))-l_2^\prime(f_2(x,y),f_0(z))\\
\label{eqn:DGLA morphism c 2}
&&-f_2(l_2(x,y),z)+f_2(x,l_2(y,z))-f_2(y,l_2(x,z))-l_3^\prime(f_0(x),f_0(y),f_0(z))=0.
\end{eqnarray}
\end{defi}
In particular, if $\huaV$ and $\huaV^\prime$ are 2-term
$L_\infty$-algebras and $f_2$ is skew-symmetric, we recover the
definition of morphisms between 2-term $L_\infty$-algebras.

If $(f_0,f_1)$ is an isomorphism of underlying complexes, we say
that $(f_0,f_1,f_2)$ is an isomorphism.

It is obvious that $2$-term sh Leibniz algebras and morphisms
between them form a category.

\section{Leibniz 2-algebras}\label{sec:leibniz2}
Leibniz 2-algebras are the categorification of Leibniz
 algebras.
Vector spaces can be categorified to $2$-vector spaces. A good
introduction for this subject is \cite{baez:2algebras}. Let $\Vect$
be the category of vector spaces.

\begin{defi}{\rm\cite{baez:2algebras}}
A $2$-vector space is a category in the category $\Vect$.
\end{defi}

Thus a $2$-vector space $C$ is a category with a vector space of
objects $C_0$ and a vector space of morphisms $C_1$, such that all
the structure maps are linear. Let $s,t:C_1\longrightarrow C_0$ be
the source and target maps respectively. Let $\cdot_\ve$ be the
composition of morphisms.

It is well known that the 2-category of 2-vector spaces is
equivalent to the 2-category of 2-term complexes of vector spaces.
Roughly speaking, given a 2-vector space $C$,
$\Ker(s)\stackrel{t}{\longrightarrow}C_0$ is a 2-term complex.
Conversely, any 2-term complex of vector spaces
$\huaV:V_1\stackrel{\dM}{\longrightarrow}V_0$ gives rise to a
2-vector space of which the set of objects is $V_0$, the set of
morphisms is $V_0\oplus V_1$, the source map $s$ is given by
$s(v+m)=v$, and the target map $t$ is given by $t(v+m)=v+\dM m$,
where $v\in V_0,~m\in V_1.$ We denote the 2-vector space associated
to the 2-term complex of vector spaces
$\huaV:V_1\stackrel{\dM}{\longrightarrow}V_0$ by $\V$:
\begin{equation}\label{eqn:V}
\V=\begin{array}{c}
\V_1:=V_0\oplus V_1\\
\vcenter{\rlap{s }}~\Big\downarrow\Big\downarrow\vcenter{\rlap{t }}\\
\V_0:=V_0.
 \end{array}\end{equation}

In this paper, we always assume that a 2-vector space is of the
above form.\vspace{3mm}

\begin{defi}
  A Leibniz $2$-algebra is a $2$-vector space $\V$ endowed with
   a bilinear functor (bracket)
  $\Courant{\cdot,\cdot}:\V\times\V\longrightarrow\V$ and a natural isomorphism $J_{x,y,z}$ for
  every $x,y,z\in\V_0$,
\begin{equation}\label{defiLeibniz21}
J_{x,y,z}:\Courant{\Courant{x,y},z}\longrightarrow
\Courant{x,\Courant{y,z}}-\Courant{y,\Courant{x,z}},
\end{equation}
such that the following Jacobiator identity is satisfied:
\begin{eqnarray}
\nonumber&&J_{\Courant{w,x},y,z}(J_{w,x,\Courant{y,z}}-\Courant{y,J_{w,x,z}})=\\\label{defiLeibniz22}
&&\Courant{J_{w,x,y},z}(J_{w,\Courant{x,y},z}-J_{x,\Courant{w,y},z})(\Courant{w,J_{x,y,z}}-J_{x,y,\Courant{w,z}}-\Courant{x,J_{w,y,z}}+J_{w,y,\Courant{x,z}}),
\end{eqnarray}
or, in terms of a diagram,
$$
\xymatrix{&\Courant{\Courant{\Courant{w,x},y},z}\ar[dr]^{J_{\Courant{w,x},y,z}}\ar[dl]_{\Courant{J_{w,x,y},z}}&\\
\Courant{\Courant{w,\Courant{x,y}},z}-\Courant{\Courant{x,\Courant{w,y}},z}\ar[dd]^{J_{w,\Courant{x,y},z}-J_{x,\Courant{w,y},z}}&&
\Courant{\Courant{w,x},\Courant{y,z}}-\Courant{y,\Courant{\Courant{w,x},z}}\ar[dd]_{J_{w,x,\Courant{y,z}}-\Courant{y,J_{w,x,z}}}\\
&&\\
P\ar[rr]^{\Courant{w,J_{x,y,z}}-J_{x,y,\Courant{w,z}}-\Courant{x,J_{w,y,z}}+J_{w,y,\Courant{x,z}}}&&Q}
$$
where $P$ and $Q$ are given by
\begin{eqnarray*}
  P&=&\Courant{w,\Courant{\Courant{x,y},z}}-\Courant{\Courant{x,y},\Courant{w,z}}-\Courant{x\Courant{\Courant{w,y},z}}+\Courant{\Courant{w,y},\Courant{x,z}},\\
  Q&=&\Courant{w,\Courant{x,\Courant{y,z}}}-\Courant{x,\Courant{w,\Courant{y,z}}}-\Courant{y,\Courant{w,\Courant{x,z}}}+\Courant{y,\Courant{x,\Courant{w,z}}}.
\end{eqnarray*}
\end{defi}
In particular, if the Jacobiator is trivial, we call a {\bf strict
Leibniz 2-algebra}; if the bilinear functor $\Courant{\cdot,\cdot}$
and the trilinear natural isomorphism $J$ are skew-symmetric, we
recover the notion of {\bf semistrict Lie 2-algebras}
\cite{baez:2algebras}.

\begin{defi}
  Let $\V$ and $\V^\prime$ be two Leibniz 2-algebras, a morphism
  from $\V$ to $\V^\prime$ consists of
  \begin{itemize}
    \item[$\bullet$] a linear functor $F$ from the underlying
    $2$-vector space of $\V$ to that of $\V^\prime$,
    \item[$\bullet$] a skewsymmetric natural transformation $$F_2(x,y):F_0(l_2(x,y))\longrightarrow l_2^\prime(F_0(x),F_0(y)),$$
  \end{itemize}
  such that
\begin{eqnarray*}
  &&(F_1J_{x,y,z})(F_2(x,\Courant{y,z})-F_2(y,\Courant{x,z}))(\Courant{F_0(x),F_2(y,z)}-\Courant{F_0(y),F_2(x,z)})\\
  &=&F_2(\Courant{x,y},z)(\Courant{F_2(x,y),F_0(z)})(J_{F_0(x),F_0(y),F_0(z)}),
\end{eqnarray*}
or in terms of diagram,
$$
\xymatrix{F_0\Courant{\Courant{x,y},z}\ar[dd]^{F_2(\Courant{x,y},z)}\ar[rr]^{F_1J_{x,y,z}}&&F_0(\Courant{x,\Courant{y,z}}-\Courant{y,\Courant{x,z}})\ar[dd]_{F_2(x,\Courant{y,z})-F_2(y,\Courant{x,z})}\\
&&\\
\Courant{F_0\Courant{x,y},F_0(z)}\ar[dd]^{\Courant{F_2(x,y),F_0(z)}}&&
\Courant{F_0(x),F_0\Courant{y,z}}-\Courant{F_0(y),F_0\Courant{x,z}}\ar[dd]_{\Courant{F_0(x),F_2(y,z)}-\Courant{F_0(y),F_2(x,z)}}\\
&&\\
\Courant{\Courant{F_0(x),F_0(y)},F_0(z)}\ar[rr]^{J_{F_0(x),F_0(y),F_0(z)}\qquad\qquad\quad}&&\Courant{F_0(x),\Courant{F_0(y),F_0(z)}}-\Courant{F_0(y),\Courant{F_0(x),F_0(z)}}.}
$$
\end{defi}

It is obvious that Leibniz algebras and morphisms between them form
a category. In the case of semistrict Lie 2-algebras, it is well
known that the category of semistrict Lie 2-algebras and the
category of 2-term $L_\infty$-algebras are equivalent \cite[Theorem
4.3.6]{baez:2algebras}. Similarly, we have

\begin{thm}\label{thm:equivalence1}
  The category of  Leibniz $2$-algebras
  and the category of $2$-term sh Leibniz algebras are equivalent.
\end{thm}
\pf  We only give a sketch on how to construct a Leibniz 2-algebra
from a 2-term sh Leibniz algebra and how to construct a 2-term sh
Leibniz algebra from a Leibniz 2-algebra. The other proof is similar
to Theorem 4.3.6 in \cite{baez:2algebras}. We omit details.

Let $(V_1\stackrel{\dM}{\longrightarrow}V_0,l_2,l_3)$ be a 2-term sh
Leibniz algebra, we  introduce a bilinear
 functor  $\Courant{\cdot,\cdot}$ on the $2$-vector space $\V$ given by \eqref{eqn:V} by
 $$
\Courant{x+m,y+n}=l_2(x,y)+l_2(x,n)+l_2(m,y)+l_2(m,\dM n).
 $$
It is straightforward to see that it does not satisfy the Leibniz
rule and the Jacobiator is given by
$$
J_{x,y,z}=\Courant{\Courant{x,y},z}+l_3(x,y,z).
$$
By Condition (f), it is not hard to see that \eqref{defiLeibniz22}
is satisfied. Thus from a 2-term sh Leibniz algebra, we can obtain a
Leibniz 2-algebra.

Conversely, given a Leibniz 2-algebra $\V$, we define $l_2$ and
$l_3$ on the 2-term complex $V_1\stackrel{\dM}{\longrightarrow}V_0$
by
\begin{itemize}
  \item[$\bullet$] $l_2(x,y)=\Courant{x,y},\quad\forall~x,y\in V_0.$
  \item[$\bullet$] $l_2(x,m)=\Courant{x,m},~l_2(m,x)=\Courant{m,x}, \quad\forall~x\in V_0,~m\in
  V_1.$
  \item[$\bullet$] $l_2(m,n)=0, \quad\forall~m,n\in V_1.$
  \item[$\bullet$] $l_3(x,y,z)=Pr_1J_{x,y,z},\quad\forall~x,y,z\in
  V_0,$ where $Pr_1:\V_1=V_0\oplus V_1\longrightarrow V_1$ is the projection.
\end{itemize}
Then one can verify that
$(V_1\stackrel{\dM}{\longrightarrow}V_0,l_2,l_3)$ is a 2-term dg
Leibniz algebra.  \qed\vspace{3mm}

\section{Omni-Lie
2-algebras}\label{sec:example}

From now on, when we say a Leibniz 2-algebra, we mean a 2-term sh
Leibniz algebra. In this section, we provide an example of Leibniz
$2$-algebras which comes from omni-Lie $2$-algebras \cite{omniLie2},
which is the categorification of Weinstein's omni-Lie
algebras.\vspace{3mm}

Let $\huaV:V_1\stackrel{\dM}{\longrightarrow}V_0$ be a complex of
vector spaces. Define $\End^0_\dM(\huaV)$ by
$$
\End^0_\dM(\huaV)\triangleq\{(A_0,A_1)\in\gl(V_0)\oplus
\gl(V_1)|A_0\circ\dM=\dM\circ A_1\},
$$
and define $\End^1(\huaV)\triangleq \End(V_0,V_1)$. There is a
differential $\delta:\End^1(\huaV)\longrightarrow \End^0_\dM(\huaV)$
given by
$$
\delta(\phi)\triangleq\phi\circ\dM+\dM\circ\phi,\quad\forall~\phi\in\End^1(\huaV),
$$
and a bracket operation $[\cdot,\cdot]$ given by the graded
commutator. More precisely,  for any $A=(A_0,A_1),B=(B_0,B_1)\in
\End^0_\dM(\huaV)$ and $\phi\in\End^1(\huaV)$, $[\cdot,\cdot]$ is
given by
\begin{eqnarray*}
  [A,B]=A\circ B-B\circ A=(A_0\circ B_0-B_0\circ A_0,A_1\circ B_1-B_1\circ
  A_1),
\end{eqnarray*}
and
\begin{equation}\label{representation}
  ~[A,\phi]=A\circ \phi-\phi\circ A=A_1\circ \phi-\phi\circ A_0.
\end{equation}
These two operations make $\End^1(\huaV)\xrightarrow{\delta}
\End^0_\dM(\huaV)$ into a 2-term DGLA (proved in
\cite{SZintegration}), which we denote by $\End(\huaV)$. It plays
the same role as $\gl(V)$ for a vector space $V$. \vspace{3mm}

Let $\frkf=(f_0,f_1,f_2)$ be an automorphism of the 2-term DGLA
$\End(\huaV)$. On the complex
$$\End(\huaV)\oplus\huaV=\End^1(\huaV)\oplus
V_1\stackrel{\delta\oplus\dM}{\longrightarrow}\End^0_\dM(\huaV)\oplus
V_0,$$ define bilinear map $l_2^\frkf$ by
\begin{equation}\left\{\begin{array}{rlll}
l_2^\frkf(A+u,B+v)&=& [A,B]+f_0(A)(v),&\mbox{in degree-$0$},\\
l_2^\frkf(A+u,\phi+m)&=&[A,\phi]+f_0(A)(m),&\mbox{in degree-$1$},\\
l_2^\frkf(\phi+m,A+u)&=&[\phi,A]+f_1(\phi)(u),&\mbox{in degree-$1$},
\end{array}\right.
\end{equation}
and define trilinear map $l_3^\frkf$ by
$$
l_3^\frkf(A+u,B+v,C+w)=f_2(A,B)(w).
$$
\begin{pro}\label{pro:example}
Let $\frkf=(f_0,f_1,f_2)$ be an automorphism of $\End(\huaV)$, then
$(\End(\huaV)\oplus\huaV,l_2^\frkf,l_3^\frkf)$ is a Leibniz
$2$-algebra.
\end{pro}
\pf We check that all the conditions in Definition
\ref{defi:2leibniz}
 are satisfied. By the fact that $\delta[A,\phi]=[A,\delta(\phi)]$
 and $f_0(A)$ commutes with $\dM$, we have
\begin{eqnarray*}
l_2^\frkf(A+u,(\delta+\dM)(\phi+m))&=&l_2^\frkf(A+u,\delta(\phi)+\dM
m)\\
&=&[A,\delta(\phi)]+f_0(A)(\dM m)\\
&=&\delta[A,\phi]+\dM f_0(A)(m)\\
&=&(\delta+\dM)l_2^\frkf(A+u,\phi+m),
\end{eqnarray*}
which implies that $(a)$ holds. Similarly, $(b)$ follows from
equalities $[\delta(\phi),A]=\delta[\phi,A]$ and
$f_0\circ\delta=\delta\circ f_1$. By the definition of $\delta$, it
is not hard to see that
$$
[\delta(\phi),\psi]=[\phi,\delta(\psi)],\quad
f_0(\delta(\phi))(n)=\delta(f_1(\phi))(n)=f_1(\phi)(\dM n).
$$
Thus we obtain $(c)$:
$$[(\delta+\dM)(\phi+m),\psi+n]=[\phi+m,(\delta+\dM)(\psi+n)].$$
It is straightforward to deduce that
\begin{eqnarray*}
&&l_2^\frkf(
A+u,l_2^\frkf(B+v,C+w))-l_2^\frkf( l_2^\frkf(A+u,B+v),C+w)-l_2^\frkf(B+v, l_2^\frkf(A+u,C+w)) \\
&=&[f_0(A),f_0(B)](w)-f_0([A,B])(w)\\
&=&\dM\circ f_2(A,B)(w).
\end{eqnarray*}
Thus we arrive at (d). $\rm(e_1)$ follows from
\begin{eqnarray*}
&&l_2^\frkf(
A+u,l_2^\frkf(B+v,\phi+m))-l_2^\frkf( l_2^\frkf(A+u,B+v),\phi+m)-l_2^\frkf(B+v, l_2^\frkf(A+u,\phi+m)) \\
&=&[f_0(A),f_0(B)](m)-f_0([A,B])(m)\\
&=& f_2(A,B)(\dM m)\\
&=&l_3^\frkf( A+u,B+v,(\delta+\dM)(\phi+m)).
\end{eqnarray*}
Similarly, $\rm(e_2)$ follows from the fact that
$$
[f_0(A),f_1(\phi)]-f_1[A,\phi]=f_2(A,\delta(\phi)),
$$
and $\rm(e_3)$ follows from the fact that
$$
[f_1(\phi),f_0(A)]-f_1[\phi,A]=f_2(\delta(\phi),A).
$$
Now we are left to show that $l_3^\frkf$ satisfies the Jacobiator
identity. This essentially follows from the fact that $\frkf$ is an
automorphism of $\End(\huaV)$. More precisely,
\begin{eqnarray*}
&&l_2^\frkf(D+x,l_3^\frkf(A+u,B+v,C+w))-l_2^\frkf(A+u,l_3^\frkf(D+x,B+v,C+w))\\
&&+l_2^\frkf(B+v,l_3^\frkf(D+x,A+u,C+w))+l_2^\frkf(l_3^\frkf(D+x,A+u,B+v),C+w)\\
&&-l_3^\frkf(l_2^\frkf(D+x,A+u),B+v,C+w)-l_3^\frkf(A+u,l_2^\frkf(D+x,B+v),C+w)\\
&&-l_3^\frkf(A+u,B+v,l_2^\frkf(D+x,C+w))+l_3^\frkf(D+x,l_2^\frkf(A+u,B+v),C+w)\\
&&+l_3^\frkf(D+x,B+v,l_2^\frkf(A+u,C+w))-l_3^\frkf(D+x,A+u,l_2^\frkf(B+v,C+w))\\
&=&f_0(D)f_2(A,B)(w)-f_0(A)f_2(D,B)(w)+f_0(B)f_2(D,A)(w)\\
&&-f_2([D,A],B)(w)-f_2(A,[D,B])(w)-f_2(A,B)f_0(D)(w)\\
&&+f_2(D,[A,B])(w)+f_2(D,B)f_0(A)(w)-f_2(D,A)f_0(B)(w)\\
&=&\Big([f_0(D),f_2(A,B)]-[f_0(A),f_2(D,B)]+[f_0(B),f_2(D,A)]\\
&&-f_2([D,A],B)+f_2([D,B],A)-f_2([A,B],D)\Big)(w)\\
&=&0.
\end{eqnarray*}
The last equality follows from the fact that $\frkf$ is an
automorphism of $\End(\huaV)$. This finishes the proof of
$(\End(\huaV)\oplus\huaV,l_2^\frkf,l_3^\frkf)$ being a Leibniz
$2$-algebra.\qed

\section{Twisted Courant algebroids}\label{sec:tca}

Hansen and Strobl introduced twisted Courant
  algebroids by closed $4$-forms in \cite{4form}, which  arise
  naturally from the study of three dimensional sigma models with
  Wess-Zumino term.

\begin{defi}{\rm\cite{4form}}
  A twisted Courant algebroid by a closed $4$-form $H$ is a vector bundle $E\longrightarrow M$, together
  with a
fiber metric $\pair{\cdot,\cdot}$ (so we can identify $E$ with
$E^*$), a bundle map $\rho: E \longrightarrow TM$ (called the
anchor), a bilinear bracket operation (Dorfman bracket)
$\Dorfman{\cdot,\cdot}$ on $\Gamma(E)$, and a closed $4$-form $H$
such that for any $e_1,e_2,e_3\in\Gamma(E)$, we have
\begin{eqnarray}
\label{nonskewsymmetric}\Dorfman{e,e}&=&\half\rho^*d\pair{e,e};\\
  \label{invariantinner} \rho(e_1)\pair{e_2,e_3}&=&\pair{\Dorfman{e_1,e_2},e_3}+\pair{e_2,\Dorfman{e_1,e_3}};\\
\label{TCcondition3}\rho^*(i_{\rho(e_1)\wedge \rho(e_2)\wedge
\rho(e_3)}H)&=&\Dorfman{e_1,\Dorfman{e_2,e_3}}-\Dorfman{\Dorfman{e_1,e_2},e_3}-\Dorfman{e_2,\Dorfman{e_1,e_3}},
\end{eqnarray}
where $d$ is the usual de Rham differential operator and
$\rho^*:T^*M\longrightarrow E^*$ is the dual map of $\rho$.
\end{defi}

We will denote a twisted Courant algebroid by
$(E,\pair{\cdot,\cdot},\Dorfman{\cdot,\cdot},\rho,H)$, which is
exactly the Courant algebroid (\cite{LWXmani, Roytenbergthesis}) if
$H=0$.

\begin{rmk} In \cite{Pontryagin class} and  \cite{regular CA}, the
authors defined  the  {\bf first Pontryagin class} for a quadratic
Lie algebroid as  an obstruction of the extension to a Courant
algebroid. In fact, it is not difficult to see that the closed
$4$-form $H$ given above is just the first Pontryagin class of the
corresponding ample Lie algebroid of a regular twisted Courant
algebroid.
\end{rmk}

Roytenberg proved that every Courant algebroid gives rise to a
$2$-term $L_\infty$-algebra in \cite{roytenbergshl}. Now  every
twisted Courant algebroid by a closed $4$-form gives rise to a
Leibniz 2-algebra.

\begin{thm}\label{thm:Leibniz2algebra}
  Every  twisted Courant algebroid by a closed $4$-form
$(E,\pair{\cdot,\cdot},\Dorfman{\cdot,\cdot},\rho,H)$ gives rise to
 a Leibniz $2$-algebra, whose degree-$1$ part is
 $\Omega^1(M)$, degree-$0$ part is $\Gamma(E)$, differential
 is  $\rho^*:\Omega^1(M)\longrightarrow\Gamma(E)$, the bilinear bracket operation $l_2$ is given by
\begin{equation}\label{eqn:l2}
 \left\{\begin{array}{rrl} l_2(e_1,e_2)&\triangleq&\Dorfman{e_1,e_2},\quad\forall ~e_1,e_2\in \Gamma(E),\\
  l_2(e,\xi)&\triangleq&L_{\rho(e)}\xi,\quad\forall ~e\in \Gamma(E),~\xi\in\Omega^1(M),\\
   l_2(\xi,e)&\triangleq&-i_{\rho(e)}d\xi,\end{array}\right.
\end{equation}
and the trilinear map $l_3^{H}$ is given by
\begin{equation}\label{eqn:l3}
  l_3^{H}(e_1,e_2,e_3)\triangleq i_{\rho(e_1)\wedge\rho(e_2)\wedge\rho(e_3)}H.
\end{equation}
\end{thm}

To prove this theorem, we need the following two lemmas.

\begin{lem}
  Let $(E,\pair{\cdot,\cdot},\Dorfman{\cdot,\cdot},\rho,H)$ be a twisted Courant algebroid by a closed $4$-form $H$, then for any $f\in C^\infty(M)$, we have
  \begin{eqnarray}
  \label{property1}\Dorfman{e_1,fe_2}&=&f\Dorfman{e_1,e_2}+\rho(e_1)(f)e_2;\\
  \label{property2}\Dorfman{fe_2,e_1}&=&f\Dorfman{e_2,e_1}-\rho(e_1)(f)e_2+\pair{e_1,e_2}\rho^*df;\\
    \label{property3}
    J_{e_1,e_2,fe_3}&=&fJ_{e_1,e_2,e_3}+\big([\rho(e_1),\rho(e_2)]-\rho\Dorfman{e_1,e_2}\big)(f)e_3.
  \end{eqnarray}
\end{lem}
\pf By \eqref{invariantinner}, we have
\begin{eqnarray*}
\rho(e_1)\pair{fe_2,e_3}&=&\pair{\Dorfman{e_1,fe_2},e_3}+\pair{fe_2,\Dorfman{e_1,e_3}}.
\end{eqnarray*}
On the other  hand, we have
\begin{eqnarray*}
\rho(e_1)\pair{fe_2,e_3}&=&\rho(e_1)(f\pair{e_2,e_3})\\
&=&\rho(e_1)(f)\pair{e_2,e_3}+f\rho(e_1)\pair{e_2,e_3}\\
&=&\pair{\rho(e_1)(f)e_2,e_3}+\pair{f\Dorfman{e_1,e_2},e_3}+\pair{fe_2,\Dorfman{e_1,e_3}}.
\end{eqnarray*}
Thus we have
\begin{eqnarray*}
\pair{\rho(e_1)(f)e_2,e_3}+\pair{f\Dorfman{e_1,e_2},e_3}=\pair{\Dorfman{e_1,fe_2},e_3}.
\end{eqnarray*}
Since the fiber metric is nondegenerate, we have
$$
\Dorfman{e_1,fe_2}=f\Dorfman{e_1,e_2}+\rho(e_1)(f)e_2.
$$

By \eqref{nonskewsymmetric}, first we have
$$
\Dorfman{e_2,e_1}+\Dorfman{e_1,e_2}=\rho^*d\pair{e_1,e_2}.
$$
Therefore, we have
$$
\Dorfman{fe_2,e_1}+\Dorfman{e_1,fe_2}=\rho^*d\pair{e_1,fe_2}=f\rho^*d\pair{e_1,e_2}+\pair{e_1,e_2}\rho^*df,
$$
which implies that
\begin{eqnarray*}
  \Dorfman{fe_2,e_1}&=&-\Dorfman{e_1,fe_2}+f(\Dorfman{e_2,e_1}+\Dorfman{e_1,e_2})+\pair{e_1,e_2}\rho^*df\\
  &=&f\Dorfman{e_2,e_1}-\rho(e_1)(f)e_2+\pair{e_1,e_2}\rho^*df.
\end{eqnarray*}
 By \eqref{property1}, we have
\begin{eqnarray*}
   J_{e_1,e_2,fe_3}&=&\Dorfman{e_1,\Dorfman{e_2,fe_3}}-\Dorfman{\Dorfman{e_1,e_2},fe_3}-\Dorfman{e_2,\Dorfman{e_1,fe_3}}\\
   &=&\Dorfman{e_1,f\Dorfman{e_2,e_3}+\rho(e_2)(f)e_3}-f\Dorfman{\Dorfman{e_1,e_2},e_3}-\rho\Dorfman{e_1,e_2}(f)e_3\\
   &&-\Dorfman{e_2,f\Dorfman{e_1,e_3}+\rho(e_1)(f)e_3}\\
   &=&f\Dorfman{e_1,\Dorfman{e_2,e_3}}+\rho(e_1)(f)\Dorfman{e_2,e_3}
+\rho(e_2)(f)\Dorfman{e_1,e_3}+\rho(e_1)\rho(e_2)(f)e_3\\&&-f\Dorfman{e_2,\Dorfman{e_1,e_3}}-\rho(e_2)(f)\Dorfman{e_1,e_3}
-\rho(e_1)(f)\Dorfman{e_2,e_3}-\rho(e_2)\rho(e_1)(f)e_3\\
&&-f\Dorfman{\Dorfman{e_1,e_2},e_3}-\rho\Dorfman{e_1,e_2}(f)e_3\\
&=&fJ_{e_1,e_2,e_3}+\big([\rho(e_1),\rho(e_2)]-\rho\Dorfman{e_1,e_2}\big)(f)e_3.
\end{eqnarray*}
The proof is completed. \qed

\begin{lem}\label{cor:invariant inner}
Let $(E,\pair{\cdot,\cdot},\Dorfman{\cdot,\cdot},\rho,H)$ be a
twisted Courant algebroid by a closed $4$-form $H$, then  we have
  \begin{eqnarray}
   \label{rhomorphism} \rho\Dorfman{e_1,e_2}&=&[\rho(e_1),\rho(e_2)],\\
   \label{rhorhostar}\rho\circ \rho^*&=&0,\\
\label{rhostarkernal1}\Dorfman{\rho^*(\xi),e}&=&\rho^*(-i_{\rho(e)}d\xi),\\
  \label{rhostarkernal2}\Dorfman{e,\rho^*(\xi)}&=&\rho^*(L_{\rho(e)}\xi).
\end{eqnarray}
\end{lem}
\pf By \eqref{TCcondition3}, we have
$J_{e_1,e_2,fe_3}=fJ_{e_1,e_2,e_3}$. Now \eqref{rhomorphism} is a
consequence of \eqref{property3}.

By \eqref{rhomorphism}, we have
\begin{eqnarray*}
\rho\Dorfman{fe_2,e_1}&=&[f\rho(e_2),\rho(e_1)]\\
&=&f[\rho(e_2),\rho(e_1)]-\rho(e_1)(f)\rho(e_2).
\end{eqnarray*}
By \eqref{property2}, we have
\begin{eqnarray*}
\rho\Dorfman{fe_2,e_1}&=&f[\rho(e_2),\rho(e_1)]-\rho(e_1)(f)\rho(e_2)+\pair{e_1,e_2}\rho\circ\rho^*df.
\end{eqnarray*}
Thus we have $\pair{e_1,e_2}\rho\circ\rho^*df=0$, which implies
\eqref{rhorhostar}.

It is not hard to deduce that
$$
\Dorfman{\rho^*(df),e}=0,\quad
\Dorfman{e,\rho^*(df)}=\rho^*(d\rho(e)(f)).
$$
Thus for any $g\in \CWM$, we have
\begin{eqnarray*}
  \Dorfman{\rho^*(fdg),e}&=&\Dorfman{f\rho^*(dg),e}\\
  &=&f\Dorfman{\rho^*(dg),e}-\rho(e)(f)\rho^*(dg)+\pair{\rho^*(dg),e}\rho^*(df)\\
  &=&-\rho(e)(f)\rho^*(dg)+\rho(e)(g)\rho^*(df)\\
  &=&\rho^*(-i_{\rho(e)}df\wedge dg)\\
  &=&\rho^*(-i_{\rho(e)}d(fdg)),
\end{eqnarray*}
and
\begin{eqnarray*}
  \Dorfman{e,\rho^*(fdg)}&=&\Dorfman{e,f\rho^*(dg)}\\
  &=&f\Dorfman{e,\rho^*(dg)}+\rho(e)(f)\rho^*(dg)\\
  &=&f\rho^*(d\rho(e)(g))+\rho(e)(f)\rho^*(dg)\\
  &=&\rho^*(L_{\rho(e)}fdg),
\end{eqnarray*}
which implies that for any $\xi\in\Omega^1(M)$, we have
$$
 \Dorfman{\rho^*(\xi),e}=\rho^*(-i_{\rho(e)}d(\xi)),
$$
and
$$
 \Dorfman{e,\rho^*(\xi)}=\rho^*(L_{\rho(e)}\xi). \qed
$$

{\bf The proof of Theorem \ref{thm:Leibniz2algebra}:} We need to
show that all the axioms in Definition \ref{defi:2leibniz} hold. By
\eqref{rhostarkernal1} and \eqref{rhostarkernal2}, it is not hard to
see that (a) and (b) hold. (c) follows from the fact that
$$
l_2(\rho^*(\xi),\eta)=l_2(\xi,\rho^*(\eta))=0.
$$
By the definition of twisted Courant algebroids and $l_3^H$, (d) is
obvious. By the definition of $l_3^H$ and \eqref{rhorhostar}, we
have
$$
l_3^H(\rho^*(\xi),e_1,e_2)=l_3^H(e_1,\rho^*(\xi),e_2)=l_3^H(e_1,e_2,\rho^*(\xi))=0.
$$
On the other hand, we have
\begin{eqnarray*}
&&l_2(e_1,l_2(e_2,\xi))-l_2(l_2(e_1,e_2),\xi)-l_2(e_2,l_2(e_1,\xi))\\&=&L_{\rho(e_1)}L_{\rho(e_2)}\xi-L_{\rho\Dorfman{e_1,e_2}}\xi-L_{\rho(e_2)}L_{\rho(e_1)}\xi\\
&=&[L_{\rho(e_1)},L_{\rho(e_2)}]\xi-L_{[\rho(e_1),\rho(e_2)]}\xi\\
&=&0,
\end{eqnarray*}
which implies that $\rm (e_1)$ holds. Similarly, it is
straightforward to see that  $\rm (e_2)$ and  $\rm (e_3)$ follow
from the formula
$$
i_{[\rho(e_1),\rho(e_2)]}d\xi=L_{\rho(e_1)}i_{\rho(e_2)}d\xi-i_{\rho(e_2)}L_{\rho(e_1)}d\xi.
$$
 At last, we need to show
that the Jacobiator identity holds. Note that $l_3^H$ is
skew-symmetric, the Jacobiator identity is equivalent to that
\begin{eqnarray*}
&&l_2(e_1,l_3^H(e_2,e_3,e_4))-l_2(e_2,l_3^H(e_1,e_3,e_4))+l_2(e_3,l_3^H(e_1,e_2,e_4))+l_2(l_3^H(e_1,e_2,e_3),e_4)\\
&&+\sum_{i<j}(-1)^{i+j}l_3^H(l_2(e_i,e_j),e_1,\dots,\widehat{e_i},\dots,\widehat{e_j},\dots,e_4)=0.
\end{eqnarray*}
Let the left hand side act on an arbitrary vector field
$X\in\frkX(M)$, we get
\begin{eqnarray*}
&&\rho(e_1)H(\rho(e_2),\rho(e_3),\rho(e_4),X)-H(\rho(e_2),\rho(e_3),\rho(e_4),[\rho(e_1),X])\\
&&-\rho(e_2)H(\rho(e_1),\rho(e_3),\rho(e_4),X)+H(\rho(e_1),\rho(e_3),\rho(e_4),[\rho(e_2),X])\\
&&+\rho(e_3)H(\rho(e_1),\rho(e_2),\rho(e_4),X)-H(\rho(e_1),\rho(e_2),\rho(e_4),[\rho(e_3),X])\\
&&-d\big(H(\rho(e_1),\rho(e_2),\rho(e_3)\big)(\rho(e_4),X)\\
&&+\sum_{i<j}(-1)^{i+j}H([\rho(e_i),\rho(e_j)],\rho(e_1),\dots,\widehat{e_i},\dots,\widehat{e_j},\dots,\rho(e_4),X),
\end{eqnarray*}
which is exactly
$$
dH(\rho(e_1),\rho(e_2),\rho(e_3),\rho(e_4),X).
$$
Since $H$ is closed $4$-form, thus $dH=0$. Therefore, $l_3^H$
satisfies the Jacobiator identity. This finishes the proof of
$(\Omega^1(M)\stackrel{\rho^*}{\longrightarrow
}\Gamma(E),l_2,l_3^{H})$ being a Leibniz $2$-algebra.
\qed\vspace{3mm}

  A twisted Courant algebroid by a closed $4$-form $H$
  $(E,\pair{\cdot,\cdot},\Dorfman{\cdot,\cdot},\rho,H)$ is said to be {\bf exact} if
  we have the following exact sequence
\begin{equation}\label{Seq:DE}
\xymatrix@C=0.5cm{0 \ar[r] & T^*M  \ar[rr]^{\rho^*} &&
               E \ar[rr]^{\rho} && TM \ar[r]  & 0.
                }
\end{equation}

By choosing an isotropic splitting $s:TM\longrightarrow E$, as
vector bundles, we have
$$E\cong\huaT\triangleq TM\oplus T^*M.$$
 We can transfer the twisted
Courant algebroid structure to $TM\oplus T^*M$. For any
$X+\xi,Y+\eta\in\frkX(M)\oplus \Omega(M)$, we have
\begin{equation}\label{rho}
\rho(X+\xi)=\rho(s(X)+\rho^*(\xi))=X,
\end{equation}
\begin{equation}\label{pair}
  \pair{X+\xi,Y+\eta}=
  \pair{s(X)+\rho^*(\xi),s(Y)+\rho^*(\eta)}=\xi(Y)+\eta(X),
\end{equation}
and
\begin{eqnarray*}
 \Dorfman{X+\xi,Y+\eta}&=&\Dorfman{s(X)+\rho^*(\xi),s(Y)+\rho^*(\eta)}\\
 &=&\Dorfman{s(X),s(Y)}+\Dorfman{s(X),\rho^*(\eta)}+\Dorfman{\rho^*(\xi),s(Y)}.
\end{eqnarray*}

By \eqref{rhomorphism} and \eqref{rhorhostar}, we have
$$
\rho\Dorfman{s(X),\rho^*(\eta)}=0,
$$
which implies that $\Dorfman{s(X),\rho^*(\eta)}\in\Omega^1(M)$. For
any $Z\in\frkX(M)$, by \eqref{invariantinner}, we have
\begin{eqnarray*}
  \Dorfman{s(X),\rho^*(\eta)}(Z)&=&\pair{\Dorfman{s(X),\rho^*(\eta)},s(Z)}\\
  &=&X\pair{\rho^*(\eta),s(Z)}-\pair{\rho^*(\eta),\Dorfman{s(X),s(Z)}}\\
  &=&X\pair{\eta,Z}-\eta([X,Z])\\
  &=&L_X\eta(Z),
\end{eqnarray*}
which implies that
\begin{equation}\label{bracket1}
\Dorfman{X,\eta}=L_X\eta.
\end{equation}
Similarly, we have
\begin{eqnarray*}
 \Dorfman{\rho^*(\xi),s(Y)}(Z)&=&\pair{\Dorfman{\rho^*(\xi),s(Y)},s(Z)}\\
 &=&\pair{-\Dorfman{s(Y),\rho^*(\xi)}+\rho^*d\pair{\rho^*(\xi),s(Y)},s(Z)}\\
  &=&-L_Y\xi(Z)+d(\xi(Y))(Z)\\
  &=&-(i_Yd\xi)(Z),
\end{eqnarray*}
which implies that
\begin{equation}\label{bracket2}
\Dorfman{\xi,Y}=-i_Yd\xi.
\end{equation}
By \eqref{rhomorphism}, we can assume that
$\Dorfman{s(X),s(Y)}-s[X,Y]=h(X,Y)$ for some
$h:\frkX(M)\times\frkX(M)\longrightarrow \Omega^1(M)$. Thus we have
\begin{equation}\label{bracket3}
\Dorfman{X,Y}=[X,Y]+h(X,Y).
\end{equation}
 It is not hard to deduce that $h\in\Omega^3(M)$. To summarize, we have

\begin{thm}\label{thm:exacthomotopyC}
 For any exact twisted Courant algebroid by a closed $4$-form $H$
 $(E,\pair{\cdot,\cdot},\Dorfman{\cdot,\cdot},\rho,H)$, as a vector
 bundle, we have $E\cong \huaT\triangleq TM\oplus T^*M$. Transfer the twisted
 Courant algebroid structure to $TM\oplus T^*M$, the anchor $\rho $
 and the fiber metric $\pair{\cdot,\cdot}$ are given by
 \eqref{rho} and \eqref{pair} respectively.
 The bracket $\Dorfman{\cdot,\cdot}$ is given by
\begin{equation}
  \Dorfman{X+\xi,Y+\eta}=[X,Y]+L_X\eta-i_Yd\xi+h(X,Y),
\end{equation}
for some $3$-form $h\in\Omega^3(M)$. We will denote this bracket
operation by $\Dorfman{\cdot,\cdot}_h$.

Consequently, any exact twisted Courant algebroid by a closed
$4$-form $H$
 $(E,\pair{\cdot,\cdot},\Dorfman{\cdot,\cdot},\rho,H)$ is isomorphic
 to the twisted Courant algebroid $(TM\oplus
 T^*M,\pair{\cdot,\cdot},\Dorfman{\cdot,\cdot}_h,\rho,dh)$, i.e. the
 closed $4$-form in an exact twisted Courant algebroid must be exact.
\end{thm}

It is well known that if we change different splittings, the 3-form
changed by an exact one. For any $B\in\Omega^2(M)$, define
$e^B:\huaT\longrightarrow\huaT$ by
$$
e^B(X+\xi)=X+\xi+i_XB,\quad\forall ~X+\xi\in\Gamma(\huaT).
$$
It is straightforward to deduce that
\begin{equation}\label{eqn:eBmorphism}
  e^B\Dorfman{X+\xi,Y+\eta}_{h+dB}=\Dorfman{e^B(X+\xi),e^B(Y+\eta)}_h.
\end{equation}
Thus we have
\begin{pro}\label{pro:isomorphicEHC}
Let $(E,\pair{\cdot,\cdot},\Dorfman{\cdot,\cdot},\rho,H)$ be an
exact twisted Courant algebroid. If we choose different splitting,
we obtain two isomorphic exact twisted Courant algebroid
$(\huaT,\pair{\cdot,\cdot},\Dorfman{\cdot,\cdot}_{h+dB},\rho,dh)$
and $(\huaT,\pair{\cdot,\cdot},\Dorfman{\cdot,\cdot}_h,\rho,dh)$.
The isomorphism is given by $e^B$. In particular, if $dB=0$, $e^B$
is an automorphism of the exact twisted Courant algebroid
$(\huaT,\pair{\cdot,\cdot},\Dorfman{\cdot,\cdot}_h,\rho,dh)$.
\end{pro}

For exact twisted Courant algebroids, since
$\rho:\huaT\longrightarrow TM$ is the projection,
$\rho^*:T^*M\longrightarrow \huaT$ is the inclusion map. Thus by
Theorem \ref{thm:Leibniz2algebra}, we obtain the following  Leibniz
$2$-algebra.

\begin{cor}\label{cor:2leibniz}
  Any exact twisted Courant algebroid $(\huaT,\pair{\cdot,\cdot},\Dorfman{\cdot,\cdot}_h,\rho,dh)$ gives rise to
 a Leibniz $2$-algebra, whose degree-$1$ part is
 $\Omega^1(M)$, degree-$0$ part is $\Gamma(\huaT)$, differential
 is the inclusion $\id:\Omega^1(M)\longrightarrow\Gamma(\huaT)$, the bilinear bracket operation $l_2^h$ is given by
\begin{equation}\label{eqn:l2exact}
 \left\{\begin{array}{rrl} l_2^h(X+\xi,Y+\eta)&=&\Dorfman{X+\xi,Y+\eta}_h,\\
  l_2^h(X+\xi,\eta)&=&\Dorfman{X+\xi,\eta}_h=L_X\eta,\\
   l_2^h(\eta,X+\xi)&=&\Dorfman{\eta,X+\xi}_h=-i_Xd\eta,\end{array}\right.
\end{equation}
and the trilinear map $l_3^{dh}$ is given by
\begin{equation}\label{eqn:l3exact}
  l_3^{dh}(X+\xi,Y+\eta,Z+\gamma)=i_{X\wedge Y\wedge Z}dh.
\end{equation}
\end{cor}

If we choose a different splitting for the exact twisted Courant
algebroid by a closed $4$-form $H$
$(E,\pair{\cdot,\cdot},\Dorfman{\cdot,\cdot},\rho,H)$, we obtain an
exact twisted Courant algebroid
$(\huaT,\pair{\cdot,\cdot},\Dorfman{\cdot,\cdot}_{h+dB},\rho,dh)$
for some $B\in\Omega^2(M)$. By Corollary \ref{cor:2leibniz}, we
obtain the Leibniz $2$-algebra
$(\Omega^1(M)\stackrel{\id}{\longrightarrow}\Gamma(\huaT),l_2^{h+dB},l_3^{dh})$.
By \eqref{eqn:eBmorphism}, we have

\begin{cor}
  $(f_0=e^B,$ $f_1=\Id,$ $f_2=0)$ is an isomorphism from the Leibniz $2$-algebra
$(\Omega^1(M)\stackrel{\id}{\longrightarrow}\Gamma(\huaT),l_2^{h+dB},l_3^{dh})$
to the Leibniz $2$-algebra
$(\Omega^1(M)\stackrel{\id}{\longrightarrow
}\Gamma(\huaT),l_2^h,l_3^{dh})$.
\end{cor}

In particular, if $dB=0$, $(f_0=e^B,$ $f_1=\Id,$ $f_2=0)$ is an
automorphism of the Leibniz $2$-algebra
$(\Omega^1(M)\stackrel{\id}{\longrightarrow
}\Gamma(\huaT),l_2^h,l_3^{dh})$. There is a more interesting
phenomenon that the Leibniz $2$-algebra
$(\Omega^1(M)\stackrel{\id}{\longrightarrow}\Gamma(\huaT),l_2^{h},l_3^{dh})$
has more automorphisms.

\begin{thm}\label{thm:automorphism}
For any $B\in \Omega^2(M)$,  $(f_0=e^B,$ $f_1=\Id,$ $f_2)$ is an
automorphism of the Leibniz $2$-algebra
$(\Omega^1(M)\stackrel{\id}{\longrightarrow
}\Gamma(\huaT),l_2^h,l_3^{dh})$, where $f_2$ is given by
$$f_2(X+\xi,Y+\eta)=i_{X\wedge Y}dB.$$
\end{thm}
\pf First it is obvious that
$$
f_0\circ \id=\id\circ f_1.
$$

 By straightforward computations, we have
\begin{eqnarray*}
  e^B\Dorfman{X+\xi,Y+\eta}_h&=&e^B\big([X,Y]+L_X\eta-i_Yd\xi+h(X,Y)\big)\\
  &=&[X,Y]+L_X\eta-i_Yd\xi+h(X,Y)+i_{[X,Y]}B,
\end{eqnarray*}
and
\begin{eqnarray*}
  \Dorfman{e^B(X+\xi),e^B(Y+\eta)}_h&=&\Dorfman{X+\xi+i_XB,Y+\eta+i_YB}_h\\
  &=&[X,Y]+L_X\eta+L_Xi_YB-i_Yd\xi-i_Ydi_XB+h(X,Y).
\end{eqnarray*}
Thus we have
\begin{eqnarray*}
\Dorfman{e^B(X+\xi),e^B(Y+\eta)}_h-e^B\Dorfman{X+\xi,Y+\eta}_h&=&L_Xi_YB-i_Ydi_XB-i_{[X,Y]}B\\
&=&L_Xi_YB-i_Ydi_XB-L_Xi_YB+i_YL_XB\\
&=&i_{X\wedge Y}dB.
\end{eqnarray*}
This shows that \eqref{eqn:DGLA morphism c 1} in Definition
\ref{defi:Leibniz morphism} holds. At last, we show that
\eqref{eqn:DGLA morphism c 2} in Definition \ref{defi:Leibniz
morphism} also holds. In fact, for any
$X+\xi,Y+\eta,Z+\gamma\in\Gamma(\huaT)$, first we have
$$
l_3^{dh}(X+\xi,Y+\eta,Z+\gamma)=l_3^{dh}(e^B(X+\xi),e^B(Y+\eta),e^B(Z+\gamma)).
$$
Thus the left hand side of \eqref{eqn:DGLA morphism c 2} is equal to
\begin{eqnarray*}
&&\Dorfman{e^B(X+\xi),f_2(Y+\eta,Z+\gamma)}_h-\Dorfman{e^B(Y+\eta),f_2(X+\xi,Z+\gamma)}_h\\&&-\Dorfman{f_2(X+\xi,Y+\eta),e^B(Z+\gamma)}_h
-f_2(\Dorfman{X+\xi,Y+\eta}_h,Z+\gamma)\\&&+f_2(X+\xi,\Dorfman{Y+\eta,Z+\gamma}_h)-f_2(Y+\eta,\Dorfman{X+\xi,Z+\gamma}_h),
\end{eqnarray*}
which is equal to
$$
L_Xi_{Y\wedge Z}dB-L_Yi_{X\wedge}ZdB+i_Zdi_{X\wedge
Y}dB-i_{[X,Y]\wedge Z}dB+i_{X\wedge[Y,Z]}dB-i_{Y\wedge[X,Z]}dB.
$$
Acting on arbitrary $W\in\frkX(M)$, we get
$$
d(dB)(X,Y,Z,W),
$$
which is zero since $d^2=0$. Thus \eqref{eqn:DGLA morphism c 2} in
Definition \ref{defi:Leibniz morphism} holds. Therefore, $(f_0=e^B,$
$f_1=\Id,$ $f_2)$, where $f_2$ is given by
$$f_2(X+\xi,Y+\eta)=i_{X\wedge Y}dB,$$
is a morphism of the Leibniz $2$-algebra
$(\Omega^1(M)\stackrel{\id}{\longrightarrow
}\Gamma(\huaT),l_2^h,l_3^{dh})$. It is an automorphism of Leibniz
$2$-algebras follows from the fact that  $(f_0=e^B,$ $f_1=\Id)$ is
an automorphism of the underlying complex. \qed

\section{Dirac structures of twisted Courant
algebroids}\label{sec:Dirac} Dirac structures of a twisted Courant
algebroid by a closed $4$-form can be defined as usual.

\begin{defi}
A Dirac structure of the twisted Courant algebroid
$(E,\pair{\cdot,\cdot},\Dorfman{\cdot,\cdot},\rho,H)$ is a maximal
isotropic subbundle $L$ such that the section space $\Gamma(L)$ is
closed under the bracket operation $\Dorfman{\cdot,\cdot}$.
\end{defi}

By \eqref{nonskewsymmetric}, the restriction of the bracket
operation $\Dorfman{\cdot,\cdot}$ on $\Gamma(L)$  is skew-symmetric.
In general, for Courant algebroids, the restriction of
$\Dorfman{\cdot,\cdot}$ on  a Dirac structure is a Lie bracket.
Denote the set $(\rho^*)^{-1} \Gamma(L)$ by $\Omega^1_L(M)$. Now we
have

\begin{thm}\label{thm:Lie2Dirac}
  Let $L$ be a Dirac structure of the twisted Courant algebroid by a
  closed $4$-form $H$
  $(E,\pair{\cdot,\cdot},\Dorfman{\cdot,\cdot},\rho,H)$. Then
 $$(\Omega^1_L(M)\stackrel{\rho^*}{\longrightarrow}\Gamma(L),l_2,l_3^{H})$$
  is a $2$-term $L_\infty$-algebra (Lie 2-algebra), in which the degree-$1$ part is $\Omega^1_L(M)\triangleq (\rho^*)^{-1} \Gamma(L)$, the degree-$0$ part is $\Gamma(L)$, $l_2$ and
  $l_3^{H}$ are given by \eqref{eqn:l2} and \eqref{eqn:l3}
  respectively.
\end{thm}
\pf First it is not hard to see that $(\Omega^1_L(M)
  \stackrel{\rho^*}{\longrightarrow}\Gamma(L),l_2,l_3^{H})$
  is a Leibniz $2$-sub-algebra of $( \Omega^1(M)\stackrel{\rho^*}{\longrightarrow
}\Gamma(\huaT),l_2,l_3^{H})$. In fact, for any $e_1,e_2\in
\Gamma(L)$, by the definition of Dirac structures, we have
$l_2(e_1,e_2)\in\Gamma(L)$. It is also obvious that for any
$e\in\Gamma(L)$ and $\xi\in\Omega^1_L(M)$, we have
$$
\rho^*l_2(e,\xi)=l_2(e,\rho^*(\xi))\in\Gamma(L),\quad
\rho^*l_2(\xi,e)=l_2(\rho^*(\xi),e)\in\Gamma(L),
$$
which implies that
$$
l_2(e,\xi)\in\Omega^1_L(M),\quad l_2(\xi,e)\in\Omega^1_L(M).
$$

For any $e_1,e_2,e_3\in\Gamma(L)$, on one hand, we have
$$
l_3^{H}(e_1,e_2,e_3)=i_{\rho(e_1)\wedge\rho(e_2)\wedge\rho(e_3)}H.
$$
On the other hand, we have
$$
\rho^*
i_{\rho(e_1)\wedge\rho(e_2)\wedge\rho(e_3)}H=\Dorfman{e_1,\Dorfman{e_2,e_3}}-\Dorfman{\Dorfman{e_1,e_2},e_3}-\Dorfman{e_2,\Dorfman{e_1,e_3}}\in\Gamma(L).
$$
Thus $l_3^{H}(e_1,e_2,e_3)\in\Omega^1_L(M)$. Therefore,
$(\Omega^1_L(M)\stackrel{\rho^*}{\longrightarrow}\Gamma(L),l_2,l_3^{H})$
  is a Leibniz $2$-sub-algebra of $(\Omega^1(M)\stackrel{\rho^*}{\longrightarrow
}\Gamma(\huaT),l_2,l_3^{H})$.

Since the Dirac structure $L$ is maximal isotropic, $l_2$ is
skew-symmetric. $l_3^{H}$ is also skew-symmetric. Thus
$(\Omega^1_L(M)\stackrel{\rho^*}{\longrightarrow}\Gamma(L),l_2,l_3^{H})$
  is a $2$-term $L_\infty$-algebra. \qed\vspace{3mm}

For a bi-vector field $\pi\in\frkX^2(M)$, let
$\pi^\sharp:T^*M\longrightarrow TM$ be the induced bundle map given
by
$$
\pi^\sharp(\xi)=i_\xi\pi,\quad\forall~\xi\in\Omega^1(M).
$$
Similar to the discussion in \cite{severa3form}, the graph of a
bundle map $\pi^\sharp$ is a Dirac structure of the exact twisted
Courant algebroid
  $(\huaT,\pair{\cdot,\cdot},\Dorfman{\cdot,\cdot}_h,\rho,dh)$ if
  and only if
\begin{equation}\label{eqn:twisted Poisson}
[\pi,\pi]=\frac{1}{2}\wedge^3\pi^\sharp h.
\end{equation}
\begin{defi}
  A bi-vector field $\pi\in\frkX^2(M)$ is called an
  $h$-twisted Poisson structure if \eqref{eqn:twisted Poisson}
  holds. $(M,\pi)$ is called an $h$-twisted Poisson manifold if
  $\pi$ is an $h$-twisted Poisson structure.
\end{defi}

For the case that the $3$-form $h$ is closed, i.e. $dh=0$, it is
discussed by $\rm\check{S}$evera and Weinstein in
\cite{severa3form}. See \cite{Int twist P, quantization of tp} for
more details. \vspace{3mm}

 One can introduce a bilinear
skew-symmetric bracket operation on the cotangent bundle of an
$h$-twisted Poisson manifold $(M,\pi)$ by
\begin{equation}\label{eqn:bracketpih}
[\xi,\eta]_{\pi,h}=L_{\pi^\sharp\xi}\eta-L_{\pi^\sharp\eta}\xi+d\pi(\eta,\xi)+i_{\pi^\sharp\xi\wedge\pi^\sharp\eta}h.
\end{equation}
Then we have
\begin{equation}\label{eqn:pimorphism}
\pi^\sharp[\xi,\eta]_{\pi,h}=[\pi^\sharp\xi,\pi^\sharp\eta],
\end{equation}
where $[\cdot,\cdot]_{\pi,h}$ is given by \eqref{eqn:bracketpih}. It
is well known that if $dh=0$, then $[\xi,\eta]_{\pi,h}$ is a Lie
bracket, consequently, $(T^*M,[\xi,\eta]_{\pi,h},\pi^\sharp)$ is a
Lie algebroid. Instead of a Lie algebroid, for an $h$-twisted
Poisson structure, we obtain

\begin{thm}\label{thm:Lie2hp}
  Associated to any $h$-twisted Poisson structure $\pi$,
  there is a  $2$-term $L_\infty$-algebra, of which the degree-$0$ part is
  $\Omega^1(M)$, the degree-$1$ part is $\Gamma(\Ker(\pi^\sharp))$, the
  differential is the inclusion $\id:\Gamma(\Ker(\pi^\sharp))\longrightarrow
  \Omega^1(M)$, $l_2$ and $l_3$ are given by
  \begin{eqnarray*}
    l_2(\xi,\eta)&=&[\xi,\eta]_{\pi,h},\quad\forall~\xi,\eta\in\Omega^1(M),\\
    l_2(\xi,u)&=&[\xi,u]_{\pi,h},\quad\forall~\xi\in\Omega^1(M),u\in\Gamma(\Ker(\pi^\sharp)),\\
    l_3(\xi,\eta,\gamma)&=&i_{\pi^\sharp\xi\wedge\pi^\sharp\eta\wedge\pi^\sharp\gamma}dh,\quad\forall~\xi,\eta,\gamma\in\Omega^1(M).
  \end{eqnarray*}
\end{thm}
\pf It is obvious that $l_2$ and $l_3$ are all skew-symmetric. For
any $\xi,\eta,\gamma\in \Omega^1(M)$, it is straightforward to
deduce that
\begin{equation}
[\xi,[\eta,\gamma]_{\pi,h}]_{\pi,h}-[[\xi,\eta]_{\pi,h},\gamma]_{\pi,h}-[\eta,[\xi,\gamma]_{\pi,h}]_{\pi,h}=i_{\pi^\sharp\xi\wedge\pi^\sharp\eta\wedge\pi^\sharp\gamma}dh.
\end{equation}
Thus we have
$$
l_3(\xi,\eta,\gamma)=[\xi,[\eta,\gamma]_{\pi,h}]_{\pi,h}-[[\xi,\eta]_{\pi,h},\gamma]_{\pi,h}-[\eta,[\xi,\gamma]_{\pi,h}]_{\pi,h}.
$$
On the other hand, by \eqref{eqn:pimorphism}, we have
\begin{eqnarray*}
   \pi^\sharp l_3(\xi,\eta,\gamma)&=&\pi^\sharp
   ([\xi,[\eta,\gamma]_{\pi,h}]_{\pi,h}-[[\xi,\eta]_{\pi,h},\gamma]_{\pi,h}-[\eta,[\xi,\gamma]_{\pi,h}]_{\pi,h})\\
   &=&[\pi^\sharp\xi,[\pi^\sharp\eta,\pi^\sharp\gamma]]-[[\pi^\sharp\xi,\pi^\sharp\eta],\pi^\sharp\gamma]-[\pi^\sharp\eta,[\pi^\sharp\xi,\pi^\sharp\gamma]]\\
   &=&0.
  \end{eqnarray*}
Therefore, we have
\begin{equation}\label{eqn:l3kernel}
l_3(\xi,\eta,\gamma)\in\Gamma(\Ker(\pi^\sharp)).
\end{equation} Now
we only need to show that the Jacobiator identity holds. For any
$\theta\in\Omega^1(M)$, by \eqref{eqn:l3kernel}, we have
 \begin{eqnarray*}
    &&l_2(\theta,l_3(\xi,\eta,\gamma))+c.p.(\theta,\xi,\eta,\gamma)-(l_3([\theta,\xi]_{\pi,h},\eta,\gamma)+c.p.(\theta,\xi,\eta,\gamma))\\
&=&L_{\pi^\sharp\theta}i_{\pi^\sharp\xi\wedge\pi^\sharp\eta\wedge\pi^\sharp\gamma}dh
+c.p.(\theta,\xi,\eta,\gamma)-(i_{\pi^\sharp[\theta,\xi]_{\pi,h}\wedge\pi^\sharp\eta\wedge\pi^\sharp\gamma}dh+c.p.(\theta,\xi,\eta,\gamma))\\
&=&L_{\pi^\sharp\theta}i_{\pi^\sharp\xi\wedge\pi^\sharp\eta\wedge\pi^\sharp\gamma}dh
+c.p.(\theta,\xi,\eta,\gamma)-(i_{[\pi^\sharp\theta,\pi^\sharp\xi]\wedge\pi^\sharp\eta\wedge\pi^\sharp\gamma}dh+c.p.(\theta,\xi,\eta,\gamma)),
  \end{eqnarray*}
where $c.p.$ means cyclic permutations. Act on an arbitrary vector
field $X\in\frkX(M)$, we get
\begin{eqnarray*}
    &&\Big(l_2(\theta,l_3(\xi,\eta,\gamma))+c.p.(\theta,\xi,\eta,\gamma)-(l_3([\theta,\xi]_{\pi,h},\eta,\gamma)+c.p.(\theta,\xi,\eta,\gamma))\Big)(X)\\
&=&\pi^\sharp\theta\big(
dh(\pi^\sharp\xi,\pi^\sharp\eta,\pi^\sharp\gamma,X)\big)-dh(\pi^\sharp\xi,\pi^\sharp\eta,\pi^\sharp\gamma,[\pi^\sharp\theta,X])
+c.p.(\theta,\xi,\eta,\gamma)\\
&&-(dh([\pi^\sharp\theta,\pi^\sharp\xi],\pi^\sharp\eta,\pi^\sharp\gamma,X)+c.p.(\theta,\xi,\eta,\gamma))\\
&=&\pi^\sharp\theta\big(
dh(\pi^\sharp\xi,\pi^\sharp\eta,\pi^\sharp\gamma,X)\big)+c.p.(\theta,\xi,\eta,\gamma)\\
&&-(dh([\pi^\sharp\theta,\pi^\sharp\xi],\pi^\sharp\eta,\pi^\sharp\gamma,X)+c.p.(\theta,\xi,\eta,\gamma,X))\\
&=&d(dh)(\pi^\sharp\theta,\pi^\sharp\xi,\pi^\sharp\eta,\pi^\sharp\gamma,X)-X(dh(\pi^\sharp\theta,\pi^\sharp\xi,\pi^\sharp\eta,\pi^\sharp\gamma))\\
&=&0.
  \end{eqnarray*}
The last equality follows from the fact $d^2=0$ and
$$
dh(\pi^\sharp\theta,\pi^\sharp\xi,\pi^\sharp\eta,\pi^\sharp\gamma)=i_{\pi^\sharp\gamma}dh(\pi^\sharp\theta,\pi^\sharp\xi,\pi^\sharp\eta)
=-i_{\pi^\sharp
dh(\pi^\sharp\theta,\pi^\sharp\xi,\pi^\sharp\eta)}\gamma=0.
$$
Therefore, $l_3$ satisfies the Jacobiator identity. This finishes
the proof of
$(\Gamma(\Ker(\pi))\stackrel{\id}{\longrightarrow}\Omega^1(M),l_2,l_3)$
being a 2-term $L_\infty$-algebra. \qed\vspace{3mm}

For an $h$-twisted Poisson structure $\pi$, the graph of
$\pi^\sharp$, which we denote by $\huaG_\pi\subset \huaT$, is a
Dirac structure. The 2-term $L_\infty$-algebra constructed in
Theorem \ref{thm:Lie2hp} is isomorphic to the 2-term
$L_\infty$-algebra constructed in Theorem \ref{thm:Lie2Dirac} for
the Dirac structure $\huaG_\pi$. More precisely, for the Dirac
structure $\huaG_\pi$, we have
$$
(\rho^*)^{-1}\huaG_\pi=\id^{-1}\huaG_\pi=\huaG_\pi\cap
T^*M=\Ker(\pi).
$$
Define $f_0:\Gamma(\huaG_\pi)\longrightarrow \Omega^1(M)$ by
$$
f_0(\pi^\sharp\xi+\xi)=\xi,
$$
and define $f_1:(\rho^*)^{-1}\huaG_\pi\longrightarrow\Ker(\pi)$ to
be the identity map. It is obvious that $f_0\circ \id=\id\circ f_1$.
Moreover, we have
\begin{eqnarray*}
f_0(\Dorfman{\pi^\sharp\xi+\xi,\pi^\sharp\eta+\eta}_h)&=&L_{\pi^\sharp\xi}\eta-L_{\pi^\sharp\eta}\xi+d\pi(\eta,\xi)+i_{\pi^\sharp\xi\wedge\pi^\sharp\eta}h\\
&=&[\xi,\eta]_{\pi,h}\\
&=&[f_0(\pi^\sharp\xi+\xi),f_0(\pi^\sharp\eta+\eta)]_{\pi,h}.
\end{eqnarray*}
Thus $(f_0,f_1)$ is an isomorphism of $2$-term $L_\infty$-algebras.

\begin{rmk} {\em
  The geometric structure underlying this $2$-term $L_\infty$-algebra is
  actually the $H$-twisted Lie algebroids introduced by Melchior
  Gr{\"u}tzmann in \cite{Grutzmann}. An $H$-twisted Lie algebroid is a quadruple $(E,[\cdot,\cdot],\rho,H)$ consists of a vector bundle $E\longrightarrow
  M$, a bundle map $\rho:E\longrightarrow TM$, a
  section $H:\wedge^3\Gamma(E)\longrightarrow \Gamma(\Ker(\rho))$,
  and a skew-symmetric bracket $[\cdot,\cdot]:\Gamma(E)\wedge\Gamma(E)\longrightarrow
  \Gamma(E)$ subject to the following axioms:
  \begin{eqnarray*}
    ~[e_1,[e_2,e_3]]+c.p.(e_1,e_2,e_3)&=&H(e_1,e_2,e_3),\\
    ~[e_1,fe_2]&=&f[e_2,e_2]+\rho(e_1)(f)e_2,\\
    DH&=&0,
  \end{eqnarray*}
  where $e_i\in\Gamma(E),~f\in \CWM$ and $DH:\wedge^4\Gamma(E)\longrightarrow\Gamma(\Ker(\rho))$ is defined by
\begin{eqnarray*}
    DH(e_1,e_2,e_3,e_4)&\triangleq&\sum_{i=1}^4(-1)^{i+1}[e_i,H(e_1,\dots,\widehat{e_i},\dots,e_4)]\\
    &&+\sum_{i<j}(-1)^{i+j}H([e_i,e_j],e_1\dots,\widehat{e_i},\dots,\widehat{e_j},\dots,e_4).
  \end{eqnarray*}
It is straightforward to see that for any $h$-twisted Poisson
structure $\pi$, $(T^*M,[\cdot,\cdot]_{\pi,h},\pi^\sharp,l_3)$ is an
$l_3$-twisted Lie algebroid.}
\end{rmk}

\end{document}